\documentclass{cernrep} 
\usepackage{texnames}
\usepackage[T1]{fontenc}
\begin{document}
\title{Unfolding in ATLAS}
 
\author{Georgios Choudalakis}

\institute{University of Chicago, on behalf of the ATLAS Collaboration}

\maketitle 

\begin{abstract}
These proceedings present the unfolding techniques used so far in ATLAS. Two representative examples are discussed in detail; one using bin-by-bin correction factors, and the other iterative unfolding.
\end{abstract}
 
\section{Introduction}
 
The distribution of any observable is distorted due to experimental limitations.
Unfolding is the procedure of estimating the ``truth-level'' spectrum, i.e., the spectrum that would be measured with an ideal detector and infinite event statistics.
A general introduction to unfolding, and details about various methods are given in other contributions to this workshop.  The focus here will be on real life examples of unfolding in ATLAS analyses.

As of early 2011, ATLAS has used two unfolding methods:
\begin{itemize}
\item[i)] bin-by-bin correction factors;
\item[ii)] the iterative method by D'Agostini \cite{DAgostini}.
\end{itemize}
One representative example will be presented from each method.  In Section~\ref{sec:bbb}, bin-by-bin correction is presented through the inclusive jet $p_T$ spectrum measurement \cite{InclusiveJet}.  In Section~\ref{sec:dagostini}, D' Agostini's iterative method\cite{DAgostini} is presented, as it was used to estimate the spectrum of charged particle multiplicity in minimum bias interactions \cite{UE}.

Both methods have drawbacks.  An insightful overview can be found in \cite{GlenCowan}, and in other contributions to this workshop.  Bin-by-bin correction has been particularly criticized for not dealing carefully with bin correlations, among other things.  ATLAS is considering methods beyond bin-by-bin in the next round of analyses where this method was used.

In searches for new physics, ATLAS does not apply any unfolding, because it is unnecessary for making a discovery, or for setting a limit to some model, or for estimating model parameters.   Unfolding can be regarded as useful when the distribution itself (or a binned version thereof) is regarded as the set of parameters of interest.

\section{Bin-by-bin correction factors}
\label{sec:bbb}

\begin{figure}
\centering\includegraphics[width=.75\linewidth]{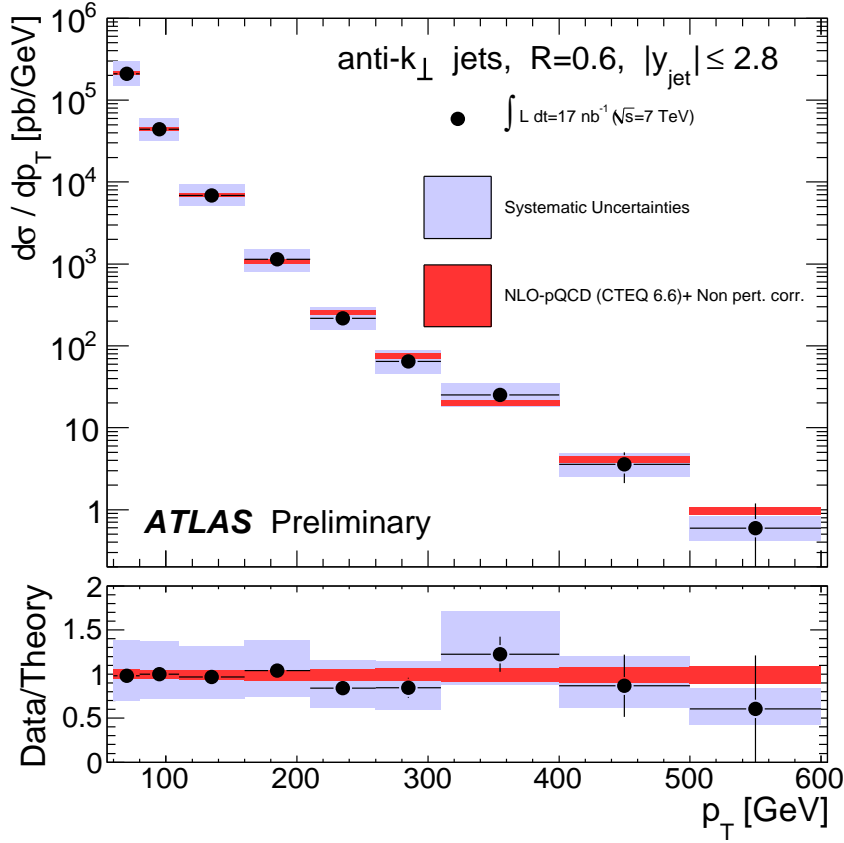}
\caption{The estimated truth-level spectrum of inclusive jet $p_T$ (filled markers) from \cite{InclusiveJet}, obtained using bin-by-bin correction factors, compared to the theoretical truth-level QCD prediction (red band).  The black error bars represent the statistical uncertainty of the estimated spectrum, and the blue band the total systematic uncertainty, which is obtained by summing in quadrature individual systematic uncertainties.  The dominant contribution comes from the jet energy scale uncertainty.  In each bin the estimated truth-level spectrum has been divided by the width of the bin and by the integrated luminosity, whose uncertainty (11\%) is not included in the blue error band.}
\label{fig:jetPt}
\end{figure}

Several ATLAS analyses have used the method of bin-by-bin correction factors \cite{InclusiveJet, InclusiveGamma, JetShape, Wjets}, mostly because of its simplicity.  The example of inclusive jet $p_T$  measurement \cite{InclusiveJet} will be discussed.  The main result of this measurement is shown in Fig.~\ref{fig:jetPt}.  In this analysis truth-level corresponds to hadron-level.

\subsection{Method description}
\label{sec:bbb:descr}

Let $T_i$ be the expected number of events in bin $i$ of the truth-level $p_T$ spectrum, which is obtained from Monte Carlo (MC).  Leading order {\sc Pythia} \cite{Pythia} QCD MC was used in the case of \cite{InclusiveJet}, where no event selection was applied.  The truth-level $p_T$ spectrum contains jets reconstructed after hadronization, applying the anti-$k_T$ clustering algorithm on stable hadrons produced after fragmentation and hadronization.  Detector simulation is not involved in the truth-level spectrum.

Let $R_i$ be the expected number of events in bin $i$ of the measured $p_T$ spectrum, after event selection which includes trigger requirements, jet reconstruction inefficiency at low $p_T$, primary vertex requirements, jet quality criteria etc.  The same {\sc Pythia} QCD MC is used as before, after ATLAS detector simulation, to obtain $R_i$.  Jets are reconstructed by applying the same anti-$k_T$ algorithm on topological clusters of energy deposited in the calorimeter \cite{JetEtMiss}.

Let $D_i$ be the actually observed number of events in bin $i$ of the measured $p_T$ spectrum.  Whereas $T_i$ and $R_i$ are both real numbers after normalizing the MC samples to the integrated luminosity of the available dataset, $D_i$ can only take integer values, because the observed events are discrete.  If it is assumed that $R_i$ is the result of an ideal simulation of all physical processes that occur at the proton collisions\footnote{Obviously this is not a good assumption when one acknowledges the possibility of new physics, but in measurements such as the one we discuss here it is presumed that what is measured is just QCD.} and of the ATLAS detector, then $D_i$ is a random integer that follows a Poisson distribution with mean $R_i$.

\begin{figure}
\centering\includegraphics[width=.75\linewidth]{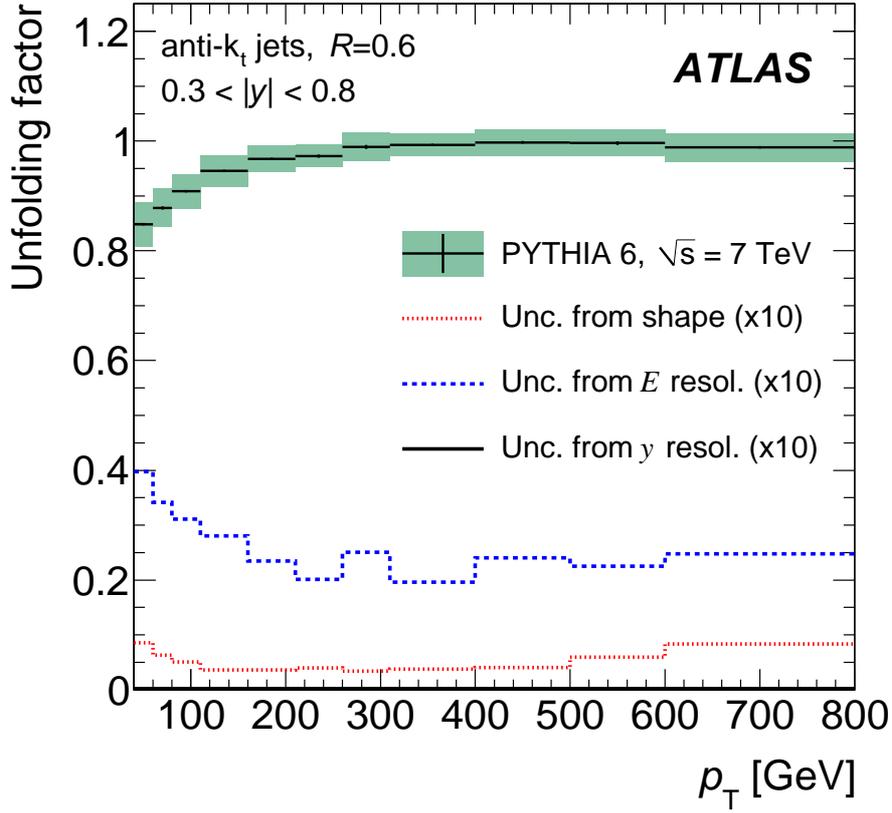}
\caption{The correction factors ($C_i$) used in \cite{InclusiveJet}.  The statistical uncertainties (black crosses) are invisibly small.  The green band represents the total systematic uncertainty, except for the part which is due to jet energy scale, which is discussed in Section~\ref{sec:JES}.}
\label{fig:corrFactors}
\end{figure}

Let 
\begin{equation}
C_i \equiv \frac{T_i}{R_i} ,
\end{equation}
 be the correction factor corresponding to bin $i$ of the observed $p_T$ spectrum.
 The correction factors used in \cite{InclusiveJet} are shown in Fig.~\ref{fig:corrFactors}.
 
The answer returned for bin $i$ of the truth-level $p_T$ spectrum after bin-by-bin correction is 
\begin{equation}
U_i \equiv C_i \cdot D_i .
\end{equation}
$U_i$ is the estimator of $T_i$.

\subsubsection{Bias}
The estimator $U_i$ has a bias that is easy to compute.  

Let's consider the possibility that the truth-level spectrum is actually $T'_i$, which may differ from the assumed $T_i$.  This could happen, for example, if sizable processes other than those included in {\sc Pythia} QCD are occurring in nature, or if the modeling of QCD by {\sc Pythia} is unrealistic.   
Let's also assume that the actual expected spectrum at detector level is $R'_i$, which may differ from $R_i$ for the above reasons, as well as due to unrealistic modeling of the detector response and of the quantities involved in event selection.  The bias of the estimator $U_i$ then is

\begin{equation}
\langle U_i - T'_i \rangle = \langle \frac{T_i}{R_i} D_i - T'_i \rangle  = \frac{T_i}{R_i} \langle D_i \rangle - T'_i  = \frac{T_i}{R_i} R'_i - T'_i  = \left(\frac{T_i}{R_i}  - \frac{T'_i}{R'_i}\right)R'_i .  \label{eq:bbb-bias}
\end{equation}

\subsection{Statistical uncertainty}

\begin{figure}
\centering\includegraphics[width=.9\linewidth]{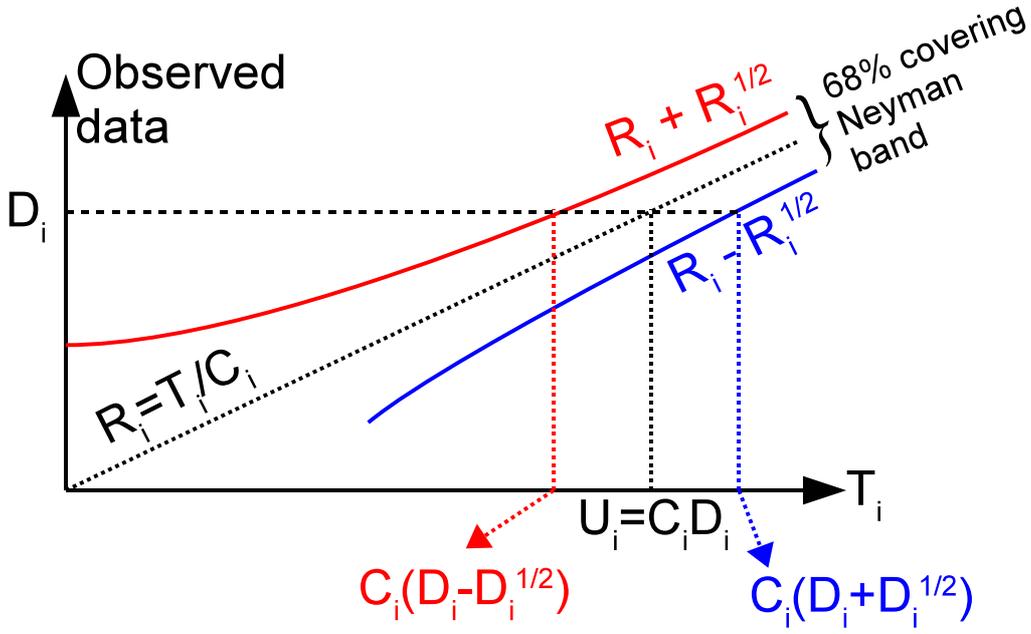}
\caption{Sketch of the Neyman construction used to correspond an observed number of data events $D_i$ to a 68\% confidence interval for $T_i$ in the bin-by-bin correction factor method.  The definitions of $D_i$, $T_i$, $R_i$, $C_i$ and $U_i$ are given in Sec.~\ref{sec:bbb:descr}.}
\label{fig:neyman}
\end{figure}

The Neyman construction shown in Fig.~\ref{fig:neyman} is effectively used to obtain a confidence interval for $T_i$, given $C_i$ and the data $D_i$.
Having observed $D_i$, the 68\% confidence interval (CI) for $U_i$ is approximately
\begin{equation}
C_i(D_i \pm \sqrt{D_i}) .
\end{equation}
This is a fair approximation when $D_i$ is large, in which case the Poisson distribution of $D_i$ with mean $R_i$ is similar to a Gaussian of mean $R_i$ and standard deviation $\sqrt{R_i}$.
Although this approximation fails in bins with few data, the same formula was used in all $p_T$ bins, so for all bins it was assumed that the statistical uncertainty of $U_i$ is symmetric and equal to
\begin{equation}
\sigma_{U_i} = C_i \sqrt{D_i} .
\end{equation}
This is the size of the black error bars in Fig.~\ref{fig:jetPt}.

\subsection{Systematic uncertainty}

The following main sources of systematic uncertainty were identified in \cite{InclusiveJet}:
\begin{itemize}
\item[i)] the correction factor $C_i$ is subject to statistical fluctuations due to finite MC event statistics; 
\item[ii)] the amount of $p_T$ smearing in detector simulation may be unrealistic;
\item[iii)] the used spectrum of $T_i$ may be unrealistic;
\item[iv)] jet energy scale uncertainty.
\end{itemize}

The following paragraphs describe how each systematic uncertainty was propagated to the final estimator of the truth-level spectrum.

\subsubsection{Finite MC event statistics}

The correction factors $C_i$ have some uncertainty due to random fluctuations of the finite MC event statistics available to determine $T_i$ and $R_i$.
When $T_i$ fluctuates above its mean, so does $R_i$, so the two are highly correlated (Fig.~\ref{fig:TRcorr}).\footnote{Though this was not done in the analysis described here, it may be worth mentioning that $T_i$ and $R_i$ could be generated separately, using different seeds for the pseudo-random numbers in the MC generator.  This would result in statistically independent estimators $T_i$ and $R_i$.  In addition to that, since $T_i$ does not involve detector simulation, it is feasible to generate many more truth-level MC events, thus estimating $T_i$ with negligible statistical uncertainty, something that unfortunately is not easy for $R_i$.}  The statistical uncertainty in $C_i$ was computed taking this correlation into account, as follows:

\begin{figure}
\centering\includegraphics[width=.25\linewidth]{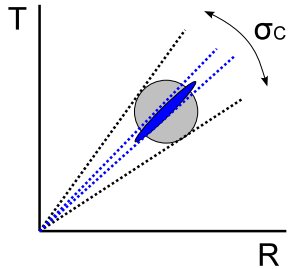}
\caption{The effect of covariance between $T_i$ and $R_i$ in the variance of the correction factor $C_i$.  The gray circle indicates the case of zero covariance, and the blue circle the case of highly positive correlation.}
\label{fig:TRcorr}
\end{figure}

The $N(R_i)$ MC evens which compose $R_i$ are separated into those that are coming from the same truth-level bin ($N(R_i \land T_i)$) and those originating from different truth-level bins ($N(R_i \land \lnot T_i)$).\footnote{The symbol $\land$ is the logical ``and'', while $\lnot$ is the logical ``not''.  So, $R_i \land \lnot T_i$ means belonging in $R_i$ {\em and not} in $T_i$.}  Similarly, the $N(T_i)$ MC events which contribute to $T_i$ are separated into those that end up in the same bin after detector simulation and event selection ($N(T_i \land R_i)$), and those that migrate to different bins ($N(T_i \land \lnot R_i)$).  The variables $N(T_i \land R_i)$ and $N(R_i \land T_i)$ are identical.  So, $C_i$ can be expressed as a function of three statistically independent random variables:
\begin{equation}
C_i = \frac{T_i}{R_i} = \frac{N(T_i \land R_i) + N(T_i \land \lnot R_i)}{N(T_i \land R_i) + N(R_i \land \lnot T_i)}.
\end{equation}
Since $C_i$ is expressed as a function of three statistically uncorrelated variables, error propagation can be used where covariance terms are zero.  Each one of the three MC event populations has standard deviation $\sqrt{N}$.

\subsubsection{Jet $p_T$ resolution uncertainty}

A relative systematic uncertainty of 15\% in jet $p_T$ resolution was assumed, based on the results of in-situ studies \cite{jerUncertainty}.

To model the effect of a different $p_T$ resolution on $C_i$, the jets in MC events were smeared by an additional amount $\alpha$, which varied from 0 to 20\% of the nominal smearing that is present in ATLAS MC.  For each amount of extra smearing, the values of $R_i$ change, while $T_i$ is not affected.  As a result each correction factor $C_i$ has a dependence on the amount of extra smearing.  It was furthermore observed that in all bins $i$ the correction factor $C_i$ increased about linearly with $\alpha$.

It is possible to increase the smearing of jet $p_T$ by adding to it a random offset of appropriate variance, but it is not possible to do the opposite, i.e., to reduce the amount of smearing that is nominally present in the ATLAS MC.  This complicates the task of determining the uncertainty on $C_i$, because the resolution uncertainty of 15\% is symmetric; the jet $p_T$ resolution could be 15\% worse or 15\% better than its nominal value.  The observation that $C_i$ depends linearly on the extra smearing justifies the assumption that, if the resolution improved, $C_i$ would decrease, linearly, at the same rate.

Therefore, the systematic uncertainty on $C_i$ due to 15\% uncertainty on jet $p_T$ resolution is determined by noting the increase of $C_i$ when resolution is made 15\% worse, and by symmetrizing this variation.  For example, if $C_i$ changes by +1\% when jet $p_T$ resolution is deteriorated by 15\%, we assign to $C_i$ a systematic uncertainty of $\pm 1\%$.

\subsubsection{Uncertainty in spectrum shape}

The correction factors $C_i$ depend on the choice of $T_i$, which affects also $R_i$.  If, for example, {\sc Pythia} QCD does not provide a realistic model of the true spectrum, that can bias $U_i$, unless $R_i$ and $T_i$ are simultaneously wrong in such a way that $T_i/R_i$ remains equal to the (unknown) actual ratio $T'_i/R'_i$ in Eq.~\ref{eq:bbb-bias}.  The use of bins quite wider than the amount of smearing makes it more likely that, even if $T_i$ is not modeled right, the ratio $T_i/R_i$ in each bin will be approximately correct.  In \cite{InclusiveJet} the bins are safely wider than jet $p_T$ resolution, and their edges are driven by experimental constraints, such as trigger thresholds.

To assess the uncertainty from possible wrong modeling of $T_i$, the MC events used to determine $C_i$ were re-weighted in multiple ways.  Their re-weighting was determined by functions smooth in jet $p_T$, chosen so as to bracket the variation observed by varying parton density functions, by including next-to-leading-order corrections to QCD, as well as the difference observed between $D_i$ and $R_i$.  For each set of re-weighted MC events both $T_i$ and $R_i$ were re-computed, and so was $C_i$ for each bin $i$.  The largest variation observed in each $C_i$ was taken as a systematic uncertainty.

\subsubsection{Jet energy scale uncertainty}
\label{sec:JES}

By far the dominant uncertainty in the final $U_i$ comes from the uncertainty in jet energy scale (JES).  All previous uncertainties, added in quadrature, amount to about 5\% of relative uncertainty in $C_i$, which is the error band shown in Fig.~\ref{fig:corrFactors}.  The rest $\sim$40\% of uncertainty in the final answer comes from the JES uncertainty, and it dominates the blue error band in Fig.~\ref{fig:jetPt}.

To propagate the JES uncertainty, the reconstructed $p_T$ of all jets in MC events is shifted by $\pm$1 standard deviation, the exact size of which is a function of jet $p_T$ and pseudo-rapidity $\eta$.  That affects $R_i$ strongly, while $T_i$ doesn't change, therefore $C_i$ varies significantly.  By applying on $D_i$ the two alternative values of $C_i$, from the positive and the negative JES shift, two extreme $U_i$ values are obtained for each bin $i$, whose distance is considered as the JES uncertainty on $U_i$.

\section{Iterative unfolding}
\label{sec:dagostini}

\begin{figure}
\centering\includegraphics[width=.6\linewidth]{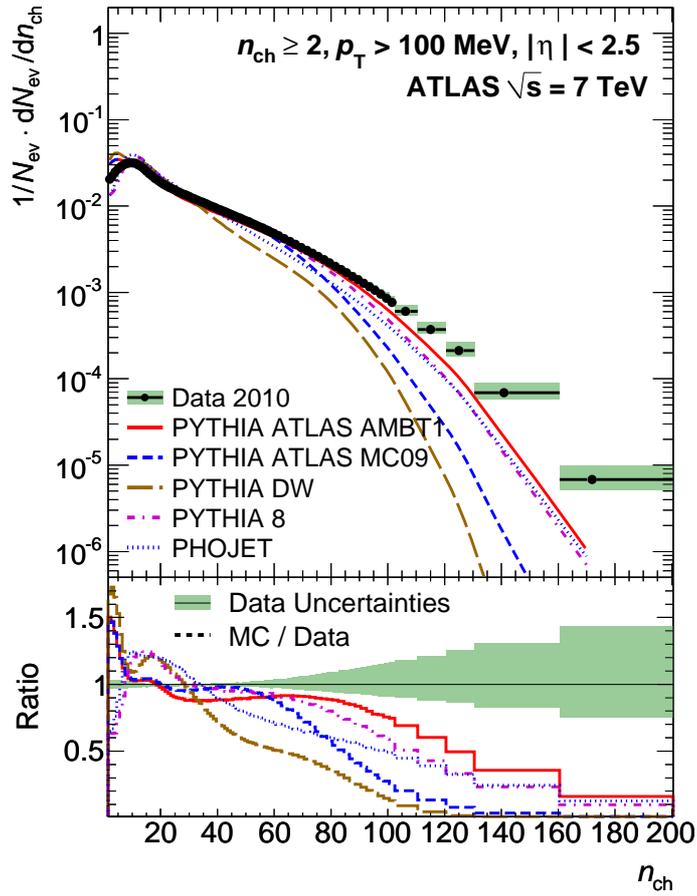}
\caption{The estimated spectrum of charged particle multiplicity (filled circles) in minimum bias $pp$ interactions, from \cite{UE}.  The statistical uncertainty is smaller than the marker size, and the asymmetric color band represents the total systematic uncertainty.  Various theoretical predictions are overlaid for comparison.}
\label{fig:nch}
\end{figure}

ATLAS used D' Agostini's iterative unfolding \cite{DAgostini} in the study of minimum bias $pp$ collisions \cite{UE}.  The example to be shown is the estimation of the truth-level distribution of the multiplicity of charged particles.  The result of this analysis is shown in Fig.~\ref{fig:nch}.

\subsection{Method description}

The full method is clearly described in the original article \cite{DAgostini} by D' Agostini.
This paragraph will make a connection between the quantities in \cite{UE} and the notation used in \cite{DAgostini}.

Let $n_{ch}$ be the number of charged particles produced in a $pp$ collision.  This is the truth-level quantity whose distribution needs to be estimated.  It corresponds to the ``cause'' $C$ mentioned in \cite{DAgostini}.

Let $n_{trk}$ be the number of reconstructed tracks in a $pp$ collision, which satisfy the selection criteria listed in \cite{UE}.  It corresponds to the ``effect'' $E$ mentioned in \cite{DAgostini}.

\begin{figure}
\centering\includegraphics[width=.5\linewidth]{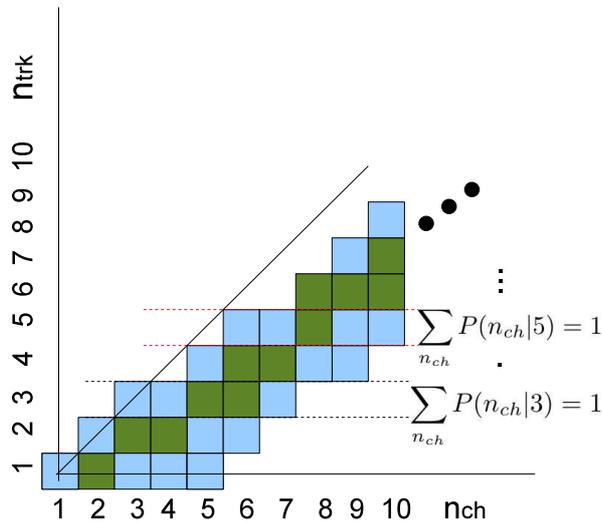}
\caption{Schematic representation of migrations matrix.  The dark green squares represent higher probability than the light blue.  Initially each matrix element equals the probability of MC events to contain $n_{ch}$ charged particles and to have $n_{trk}$ reconstructed tracks that satisfy the criteria listed in \cite{UE}.
Then, the elements of each row, corresponding to a fixed $n_{trk}$, are normalized to have sum 1.}
\label{fig:migrations}
\end{figure}

The reconstructed tracks are typically fewer than the actual charged particles, due to tracking inefficiency, therefore typically $n_{trk} \le n_{ch}$.  Therefore the migrations matrix is highly non-diagonal, and schematically looks like Fig.~\ref{fig:migrations}.

Re-writing the basic formulas from \cite{DAgostini}, substituting $C \to n_{ch}$ and $E \to n_{trk}$, we get
\begin{eqnarray}
\hat{N}(n_{ch}) &=& \frac{1}{\epsilon(n_{ch})} \sum_{n_{trk} \ge 2} N(n_{trk})P(n_{ch}|n_{trk}) \hspace{1cm} \epsilon(n_{ch}) \neq 0 , \label{eq:Nhat} \\
P(n_{ch} | n_{trk}) &=& \frac{P(n_{trk}|n_{ch})P_0(n_{ch})}{\sum_{n_{ch} \ge 1} P(n_{trk}|n_{ch}) P_0(n_{ch})} ,
\end{eqnarray}
where the efficiency $\epsilon(n_{ch})$ corresponds to the probability of reconstructing at least two tracks, a requirement related to having a reliable primary vertex reconstruction, for a given number of charged particles: 
\begin{equation}
\epsilon(n_{ch}) = P(n_{trk} \ge 2 | n_{ch}) . \label{eq:epsilon}
\end{equation}
The term $P_0(n_{ch})$ is an arbitrary initial distribution for the truth-level quantity $n_{ch}$.
The symbol $N(n_{trk})$ denotes the population of events where $n_{trk}$ tracks were reconstructed, and $\hat{N}(n_{ch})$ is the estimator of the population of events with $n_{ch}$ charged particles at truth-level.

\subsubsection{Initial distribution and iterations}

In \cite{UE}, the initial distribution was defined to be the $n_{ch}$ spectrum predicted by {\sc Pythia} minimum bias MC.  The reason is that the {\sc Pythia} prediction has been tuned to data from various past experiments, so it is a reasonable starting point.

In iterative unfolding the number of iterations is decided arbitrarily.  Too many iterations result in bin-by-bin fluctuations in the unfolded spectrum, similar to what one may get from simple migration matrix inversion \cite{GlenCowan}.  Too few iterations increase too much the influence of the initial distribution on the final answer.

In \cite{UE}, a convergence criterion was defined to determine when to stop iterating.  The criterion was
\begin{equation}
\frac{\chi^2}{N_{bins}} < 1 ,  \label{eq:chi}
\end{equation}
where 
\begin{equation}
\chi^2 \equiv \sum_{i=1}^{N_{bins}} \left( \frac{n_{ch}^{i,current} - n_{ch}^{i,previous}}{\sqrt{n_{ch}^{i,previous}}} \right)^2 .
\end{equation}
Namely, iterations continued until the latest unfolded spectrum ($n_{ch}^{current}$) remained statistically consistent with the spectrum from the previous iteration ($n_{ch}^{previous}$).  It was found that 4 iterations were enough to meet this convergence criterion.

\subsubsection{The term $\epsilon(n_{ch})$}

In principle one should extract $\epsilon(n_{ch})$ defined in Eq.~\ref{eq:epsilon}, directly from the MC events used to populate the migrations matrix (Fig.~\ref{fig:migrations}).  However, a decision was made in \cite{UE} to use instead a parametric approximation of $\epsilon(n_{ch})$.

Making the simplification that each charged particle has the same ``average effective'' probability $\epsilon_{eff}$ of being reconstructed as a track, the probability of having at least two reconstructed tracks is given by
\begin{equation}
  f(n_{ch}) = 1 - (1-\epsilon_{eff})^{n_{ch}} - n_{ch}(1-\epsilon_{eff})^{(n_{ch}-1)}\epsilon_{eff} . \label{eq:f}
\end{equation}
The unknown parameter $\epsilon_{eff}$ was adjusted so as make $f(2)$ equal to the $\epsilon(n_{ch}=2)$ obtained from MC.  The resulting value for $\epsilon_{eff}$ is within 4\% from the average probability of track reconstruction that is determined from MC simulation, which indicates that $f(n_{ch})$ matches well the MC-driven $\epsilon(n_{ch})$ even for $n_{ch} > 2$.

After adjusting $\epsilon_{eff}$ as described, the quantity $f(n_{ch})$ from Eq.~\ref{eq:f} substitutes $\epsilon(n_{ch})$ in Eq.~\ref{eq:Nhat}.
Practically this efficiency becomes $\simeq 1$ for $n_{ch} > 4$, and that is true regardless of using $f(n_{ch})$ or $\epsilon(n_{ch})$.

\subsection{Statistical uncertainty}

In Eq.~\ref{eq:Nhat}, the estimator $\hat{N}(n_{ch})$ depends on the measured $N(n_{trk})$, which are the result of independent Poisson fluctuations.  Simple error propagation would lead to the following standard deviation for $\hat{N}(n_{ch})$:
\begin{equation}
\sigma_{\hat{N}(n_{ch})} = \sqrt{\sum \left( \frac{1}{\epsilon(n_{ch})} P(n_{ch}|n_{trk}) \right)^2 N(n_{trk})} .
\end{equation}

The way statistical uncertainty was actually calculated in \cite{UE} was
\begin{equation}
\sigma_{\hat{N}(n_{ch})} = \sqrt{\hat{N}(n_{ch})}.
\end{equation}

Either way, the statistics in all bins of $n_{trk}$ are high enough to make the statistical uncertainty negligible.  In Fig.~\ref{fig:nch}, the statistical error bars are invisible.

\subsection{Systematic uncertainty}

The following main sources of systematic uncertainty will be discussed:
\begin{itemize}
\item[i)] The choice of initial distribution $P_0(n_{ch})$;
\item[ii)] The uncertainty in track reconstruction efficiency;
\item[iii)] The uncertainty in MC spectrum.
\end{itemize}

\subsubsection{Choice of initial distribution}

The stability of the answer under different choices of initial distribution $P_0(n_{ch})$ was tested by assuming a ``flat'' initial distribution $P_0(n_{ch}) = 1$, and repeating the iterative unfolding procedure.  This choice is obviously physically absurd; its purpose was only to show that even under extreme choices of $P_0(n_{ch})$ the answer $\hat{N}(n_{ch})$ doesn't change much.

Starting from a flat initial distribution, the number of iterations required to converge (Eq.~\ref{eq:chi}) increased from 4 to 7.
The final answer changed by less than 2\% in all bins of $n_{ch}$, which was taken as a systematic uncertainty in $\hat{N}(n_{ch})$.

\subsubsection{Track reconstruction efficiency uncertainty}
\label{sec:effUncertainty}

The main effect this unfolding is correcting is the inefficiency of tracking.  This inefficiency is reflected in the probabilities of Eq.~\ref{eq:Nhat}, and is obtained from MC simulation.  If tracking inefficiency in MC is wrong, so is the obtained spectrum after unfolding.

\begin{figure}
\centering\includegraphics[width=0.5\textwidth]{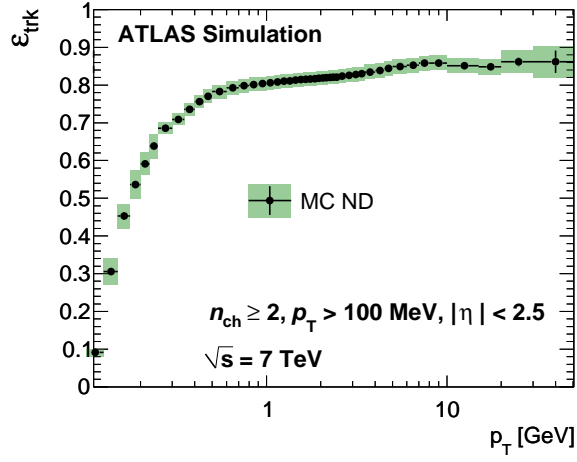}
\caption{Track reconstruction efficiency $\epsilon_{trk}$ in ATLAS simulation.  The error band represents its systematic uncertainty.}
\label{fig:trkEff}
\end{figure}

Fig.~\ref{fig:trkEff} shows the track reconstruction efficiency ($\epsilon_{trk}$) in ATLAS simulation.

To propagate the uncertainty of $\epsilon_{trk}$ into $\hat{N}(n_{ch})$, the natural thing to do would be to shift systematically $\epsilon_{trk}$, thus changing $P(n_{ch}|n_{trk})$, and see how much $\hat{N}(n_{ch})$ would change.  Instead, what was done in \cite{UE} was to keep the migration probabilities fixed, and modify the data ($N(ch_{trk})$) on which iterative unfolding was applied.  The way in which the data were modified is described next.

Assume an event in data has $n_{trk}$ tracks.  Take one of these tracks.  Its $p_T$ corresponds to some efficiency $\epsilon_{trk}$ (Fig.~\ref{fig:trkEff}).  For the sake of clarity, let's say it corresponds to $\epsilon_{trk} = 0.80 \pm 0.05$.  This $\epsilon_{trk}$ gets reduced by 1 standard deviation, so it is brought down to 0.75.  For this reduced $\epsilon_{trk}$, the expected number of tracks is $\frac{1}{0.80}\times 0.75 \simeq 0.94$.  The track is then randomly kept, with probability 0.94, or discarded, with probability 0.06.  This procedure of efficiency reduction and random removal is repeated for all $n_{trk}$ tracks of the event.  In the end, the event is left with $n'_{trk}$, where $n'_{trk} \le n_{trk}$.

The above procedure is repeated for all data events, reducing $n_{trk}$ to $n'_{trk}$ in each event.  Then, the distribution $N(n'_{trk})$ is unfolded instead of $N(n_{trk})$, which results in $\hat{N}'(n_{ch})$ instead of $\hat{N}(n_{ch})$.

The above procedure could only remove tracks, not create any.  However, the $\epsilon_{trk}$ uncertainty is symmetric, which means that the actual $\epsilon_{trk}$ could be also greater than its nominal value.  For this reason, the difference between $\hat{N}(n_{ch})$ and $\hat{N}'(n_{ch})$ is symmetrized, and used as a systematic uncertainty in $\hat{N}(n_{ch})$.  That means, for example, that if in a bin of $n_{ch}$ the $\hat{N}'(n_{ch})$ was 5\% greater than $\hat{N}(n_{ch})$, the uncertainty is set to $\pm 5\%$.

\subsubsection{Uncertainty due to spectrum shape}

\begin{figure}
\centering\includegraphics[width=0.5\textwidth]{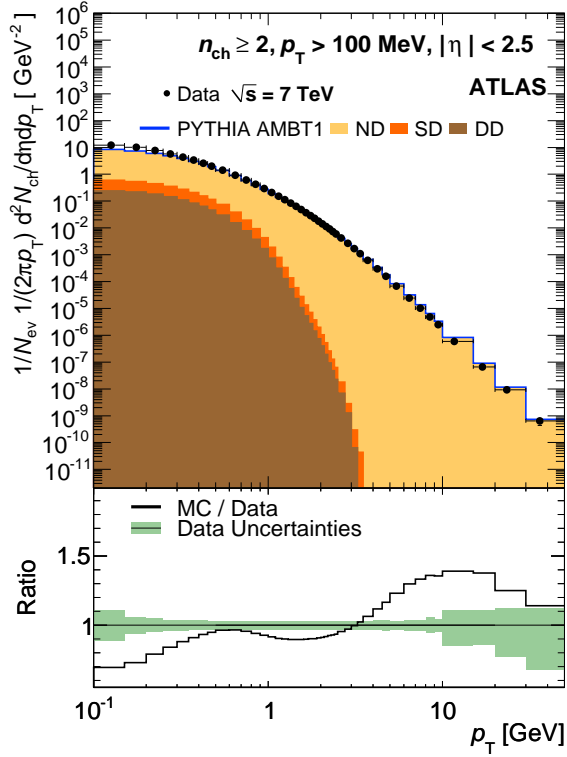}
\caption{The observed spectrum of $p_T^{trk}$, compared to the spectrum expected from MC simulation.}
\label{fig:trkPt}
\end{figure}

The observed spectrum of track transverse momentum ($p_T^{trk}$) disagrees with the MC prediction after full ATLAS detector simulation, as shown in Fig.~\ref{fig:trkPt}.  This discrepancy is related to the unfolding from $n_{trk}$ to $n_{ch}$, because $\epsilon_{trk}$ is a function of $p_T^{trk}$ (Fig.~\ref{fig:trkEff}).  If the $p_T^{trk}$ is not realistically modeled, neither is $\epsilon_{trk}$.

The way this was treated was the following:  In each bin of $n_{trk}$, the mean $\epsilon_{trk}$ was found by looping through all data events in the bin, and corresponding each observed $p_T^{trk}$ to the value of $\epsilon_{trk}$ obtained from MC (Fig.~\ref{fig:trkEff}).  The same was then done for MC events in bins of $n_{trk}$, again corresponding the $p_T^{trk}$ in MC events to values of $\epsilon_{trk}$ from Fig.~\ref{fig:trkEff}.  In each bin of $n_{trk}$, the average track reconstruction efficiency $\langle \epsilon_{trk} \rangle$ from data was compared to the same quantity from MC.  The same $p_T^{trk} \to \epsilon_{trk}$ correspondence was used for both data and MC tracks, therefore the difference in the resulting $\langle \epsilon_{trk} \rangle$ is due to the different $p_T^{trk}$ distributions.  

In each bin of $n_{trk}$, if $\langle \epsilon_{trk} \rangle$ is larger in data than in MC, then the efficiency in data gets reduced by the observed difference, in the same stochastic way described in Section~\ref{sec:effUncertainty}.  This results in a different number of tracks $n'_{trk} \le n_{trk}$ for each event in the data.  The iterative unfolding is then applied to $N(n'_{trk})$, and a different estimator of the truth-level spectrum is obtained ($\hat{N}'(n_{ch})$).  The difference between the nominal $\hat{N}(n_{ch})$ and $\hat{N}'(n_{ch})$ is found, and is used as a one-sided systematic uncertainty in $\hat{N}(n_{ch})$.

In $n_{trk}$ bins where the $\langle \epsilon_{trk} \rangle$ in data is smaller than in MC, one would ideally wish to increase the $\epsilon_{trk}$ of the data, but it is not possible to create tracks, as explained in Section~\ref{sec:effUncertainty}.  Instead, the $\epsilon_{trk}$ of data is {\em reduced} by the observed difference, as if the data had greater $\langle \epsilon_{trk} \rangle$ than the MC.  This results in a different data spectrum $N(n'_{trk})$, which after unfolding results in a different estimator $\hat{N}'(n_{ch})$.  The difference between the nominal $\hat{N}(n_{ch})$ and $\hat{N}'(n_{ch})$ is found, and instead of using it directly as a one-sided systematic uncertainty in $\hat{N}(n_{ch})$, we use its opposite, to take into account the fact that the data $\epsilon_{trk}$ was reduced instead of increased.

The above procedure results in an asymmetric systematic uncertainty.  The upper and lower uncertainty are separately added in quadrature with the other systematic uncertainties, which are symmetric.  This results in two unequal total systematic uncertainties, one suggesting that $N(n_{ch})$ could be above and the other below its nominal value.  This asymmetric systematic uncertainty appears as a colored band in Fig.~\ref{fig:nch}.

\section{Concluding remarks}

Two examples were shown of how unfolding has been used in ATLAS.  They were chosen to be representative of different cases; one is using the bin-by-bin factors to correct the spectrum of a continuous observable (jet $p_T$), whereas the other uses iterative unfolding to estimate the truth-level distribution of a discrete variable $(n_{ch})$.  The systematic uncertainties are quite different, as one analysis deals mainly with energy smearing, and the other with tracking inefficiency.

As of the time of this workshop, ATLAS has used extensively bin-by-bin correction factors, and in some cases iterative unfolding.  More methods are being considered for future iterations of some of these analyses.

Unfolding has been used only in analyses where the goal was to estimate a truth-level distribution.  Unfolding has been deliberately avoided in searches for new physics, where bias in bins with low statistics can not be afforded, where it can not be assumed that the data are consistent with the MC prediction as is silently assumed in some stages of unfolding, and where Poisson-distributed data are simpler to evaluate than estimators resulting from unfolding procedures after a series of arbitrary regularization choices.  There are several unfolding methods, in some of which anomalies due to new physics could even be reduced, whereas the observed data are unique.  Any inference is possible using directly the data, without unfolding, and a theoretical prediction that either includes full detector simulation, or at least an approximation of it that amounts to the inverse of unfolding, namely folding, which can be done with no need for regularization.  The only task for which unfolding is strictly needed is the estimation of a truth-level spectrum, whose later use to make statistical inferences becomes complicated by non-Poissonian statistics, biases that are hard to estimate, and bin-to-bin correlations.

In none of the analyses where unfolding was used was the full covariance matrix provided.  The latter would be necessary to correctly compare a truth-level theoretical prediction to the result of unfolding.  In most cases the comparison between the result of unfolding and the truth-level Standard Model prediction is made qualitatively, avoiding to provide a $p$-value that would require proper use of the covariances between bins.  When a more quantitative comparison is attempted, like in \cite{JetShape}, a $\chi^2$ is used only as a metric to determine if one theory agrees with the data more than another, but not as a test statistic to compute a $p$-value.

\section*{Acknowledgements}
I thank the organizers of PHYSTAT2011 for the opportunity to have this discussion, and my ATLAS collaborators for their feedback in preparation for this workshop.

\end{document}